%% file: Proceedings_v5.tex
\documentclass[12pt]{article}
\usepackage{epsfig}
\usepackage{graphicx}

\input econfmacros-csqcd4.tex
\def\Title#1{\begin{center} {\Large {\bf #1} } \end{center}}

\newcommand{\sym}{${\mathcal N}=4$}
\newcommand{\symm}{${\mathcal N}=4$ SYM}

\newcommand{\bea}{\begin{eqnarray}}
\newcommand{\beal}[1]{\begin{eqnarray}\label{#1}}
\newcommand{\eea}{\end{eqnarray}} 
\newcommand{\be}{\begin{equation}} 
\newcommand{\eb}{\end{equation}} 
\newcommand{\bel}[1]{\begin{equation}\label{#1}}
\newcommand{\rf}[1]{Eq.~(\ref{#1})}

\newcommand{\bit}{\begin{itemize}}
\newcommand{\eit}{\end{itemize}}
\newcommand{\ben}{\begin{enumerate}}
\newcommand{\een}{\end{enumerate}}

\begin{document}

\Title{Hydrodynamization Physics from Holography}
% and Supernovae}

\smallskip
% \bigskip

%+\addcontentsline{toc}{chapter}{{\it D. Blaschke}}
%+\label{BlaschkeDavid}

\begin{raggedright}

{
Jakub Jankowski\\
\bigskip
{\it \small
Institute of Physics, Jagiellonian University, ul. {\L}ojasiewicza 11, 30-348 Krak\'ow, Poland\\
}}
\end{raggedright}

\section{Introduction}

The quest for a better understanding of the properties of nuclear matter under
extreme conditions (such as those created in relativistic heavy ion
collisions) has led to a 
number of theoretical challenges. One of them is to explain the success of 
the hydrodynamic description of quark-gluon plasma (QGP) evolution at time scales of
order $\tau\sim0.5-1$~fm/c~\cite{Teaney:2000cw,Kolb:2003dz}. Immediately
following the nuclear collision the system is very far from equilibrium, yet rapidly reaches a
state where hydrodynamics appears to apply. This process is fast in the sense
that it occurs on a time scale less or equal the 
inverse local temperature, i.e. $\tau_HT_H\leq 1$. 

The fact that the hydrodynamic description works well with a very low value of
the shear viscosity to entropy ratio ($\eta/s$) and vanishing bulk viscosity
suggests that the plasma state may be viewed effectively as a strongly
coupled, conformal fluid. This has motivated studies aimed at understanding
how such a hydrodynamic description could emerge in a case which is amenable
to theoretical studies -- the \sym\ supersymmetric Yang-Mills theory (SYM). In
this example one has the option to apply gauge/gravity duality
\cite{Maldacena:1997re} to model the approach to equilibrium. Such studies
were initiated in~\cite{Janik:2005zt}, where the case of Bjorken flow
(discussed in more detail below) was considered. Janik and Peschanski showed
in that context that the hydrodynamic description does indeed emerge at late
proper times. This work was followed by numerous articles devoted to the
description of hydrodynamic states in the context of gauge/gravity duality. In
particular, second order transport coefficients were
computed~\cite{Heller:2007qt,Bhattacharyya:2008ji}, and many new insights into
the meaning of relativistic hydrodynamics were
gained~\cite{Baier:2007ix,Romatschke:2009im,Heller:2013fn,Heller:2014wfa,Heller:2015dha}.

To describe far from equilibrium states of \symm\ it is necessary to resort to
numerical calculations. A decisive step opening this field of research was
made by Chesler and Yaffe \cite{Chesler:2008hg} who devised a very effective
numerical scheme for solving Einstein equations in asymptotically AdS spaces
based on characteristic evolution. This was soon applied 
to the case of Bjorken flow \cite{Chesler:2009cy}, which is a particularly
attractive setting, since it was an important model used to understand basic
features of QGP evolution (such as entropy production), and at the same time is
simple enough to implement easily in the context of gauge/gravity duality.
The results described here~\cite{Jankowski:2014lna} follow from a modification
of this scheme. A different method was used in
\cite{Heller:2011ju,Heller:2012je}, where some basic features of the approach
to hydrodynamics were studied. Most importantly, it was found there that the
system reaches the hydrodynamic regime quickly (in the sense described
earlier). Another physically important 
conclusion from these papers (earlier noted in \cite{Chesler:2009cy}) was that
hydrodynamics works very well already at a time when pressure gradients are
large. Thus the process of reaching this stage of evolution is often referred
to as \emph{hydrodynamization} instead of thermalization.

The numerical studies \cite{Heller:2011ju,Heller:2012je} used a different numerical scheme from
\cite{Chesler:2008hg,Chesler:2009cy}, which was however limited by the fact
that only a few (29) consistent
initial states were known. The study \cite{Jankowski:2014lna} reported here 
adapted the approach of \cite{Chesler:2009cy} in a way which allows an arbitrary
number of initial conditions to be analysed, making it possible to look for generic
features. In our work we looked at 600 initial conditions  randomly generated
on the gravity side of the duality. In the context of gauge/gravity duality
there is a natural characteristic of 
the initial state called the initial entropy \cite{Heller:2011ju} (to be
defined 
precisely below). Observables such as
hydrodynamization time depend in particular on this quantity. 

In this note we will
focus on two natural questions
\begin{itemize}
\item Is hydrodynamization generically a fast process? 
\item Are there any universal physical characteristics of hydrodynamization? 
\end{itemize}
The answer to the first question seems to be positive and confirms previous
investigations \cite{Heller:2011ju,Heller:2012je};
regardless of the values of the initial entropy chosen the system reaches hydrodynamic
description on time scales of the order of the inverse local temperature. The second 
question is more subtle. In the sample of initial data analysed in our study
the hydrodynamization time appears not to be correlated with initial entropy (in
contrast to \cite{Heller:2011ju}, where such a correlation seemed to be
present on the basis of a smaller sample of initial 
states). However, when looking at the energy density at the time 
when hydrodynamic evolution starts, there appears to be a linear correlation
with initial entropy.

One should also mention the work \cite{Heller:2012km,Heller:2013oxa} where the
process of isotropization was considered 
in a similar spirit. Recently a technically different (closer to the original
formulation of the problem \cite{Chesler:2008hg}) but conceptually similar
studies  appeared in 
\cite{Bellantuono:2015hxa}. 

%xxx: maybe move conclusions to summary?
%xxx: what of entropy production - prominent in the article

%%%%%%%%%%%%%%%%%%%%%%%%%%%%%%%%%%%%%%%%%%%%%%%%%%%%%%%%%%%%%%%%%%%%%%%%

\section{Supersymmetric plasma and gravity}

As discussed in the introduction, \symm\ is amenable to quantitative studies
in the 
strongly coupled limit. In the absence of methods which could be used in QCD
for the study of strongly coupled, real time dynamics, this theory has become
a focus of much attention. This theory shares some features of QCD (especially at
high temperature), but is rather different from it. It contains, apart from
gluons, $6$ real massless scalars and $4$ Majorana massless 
fermions, all in the adjoint representation of the $U(N)$ gauge group. The theory is 
known to be conformal even at the quantum level. As mentioned in the introduction, we
work within the AdS/CFT correspondence \cite{Maldacena:1997re} which becomes
an effective computational tool in the 't Hooft limit $N\rightarrow\infty$ and 
$\lambda=g_{\rm YM}^2 N\rightarrow\infty$, where quantum and stringy
corrections on the gravity side can be neglected. 

There are important similarities and differences between SYM and QCD which
one has to keep in mind. Among the similarities are the existence of the
deconfined phase. Also, in the perturbative regime at $T>0$, both theories
have been shown to behave similarly, with the difference coming mostly from
the different number of degrees of freedom \cite{Czajka:2012gq}.  The most
crucial difference is that \symm\ has a vanishing beta function, which implies that it is
not confining, has no finite temperature phase transition and has an exactly
conformal equations of state.

The modeling of nuclear collisions is an extremely complex task. To reduce the
complexity of the problem we adopt strong symmetry assumptions 
introduced by Bjorken \cite{Bjorken:1982qr} for the description of matter
following a heavy ion collision.  The dynamics of the system is assumed to be independent
of boosts along the longitudinal (collision) axis and is independent of
transverse coordinates. In proper time-rapidity coordinates $ t = \tau \cosh
y$~, $z = \tau \sinh y$ this reduces to the statement that observables
dependent only on proper time $\tau$. This approximation becomes exact in the
limit of an infinite energy collision of infinitely large nuclei. 

Two physical quantities of interest to us will be the energy momentum tensor
and entropy. In the present circumstances first one takes the form
\begin{equation}
T_{\mu\nu}= {\rm Diag}\{\epsilon(\tau),p_L(\tau),p_T(\tau),p_T(\tau)\}~,
\label{eq:emtensor}
\end{equation}
where $\epsilon(\tau) = p_L(\tau)+2p_T(\tau)$ as required by conformal
symmetry. The energy density defines  
local effective temperature by the relation
\begin{equation}
\epsilon(\tau) = \frac{3}{8} N^2\pi^2T(\tau)^4~.
\label{eq:efftemp}
\end{equation}
This is the temperature of an equilibrium system with the same 
energy density. It can be shown that for late times the dynamics 
is governed by the equations of hydrodynamics; up to third order the effective
temperature
follows~\cite{Janik:2005zt,Janik:2006ft,Heller:2007qt,Booth:2009ct}   
\begin{eqnarray}
  \label{eq:Th}
  T(\tau) &=& \frac{\Lambda}{\left( \Lambda \tau \right)^{1/3}} \Big\{
  1 - \frac{1}{6 \pi \left( \Lambda \tau \right)^{2/3}} + \frac{-1 + \log{2}}{36
    \pi^{2} \left( \Lambda \tau \right)^{4/3}} +\\ \nonumber &+& \frac{-21 +
    2\pi^{2} + 51 \log{2} - 24 \log^{2}{2}}{1944 \pi^{3} \left(\Lambda
    \tau\right)^{2}} \Big\} \, .
\end{eqnarray}
The energy scale $\Lambda$ appearing in \rf{eq:Th} 
depends on the initial conditions chosen and it is the only 
trace of initial state information contained in the hydrodynamic expansion. It
also sets the scale for the energy density at hydrodynamization.

The calculation strategy follows usual lore of holography \cite{Janik:2010we}. 
States in the boundary theory correspond to asymptotically AdS geometries 
in the bulk. For example an equilibrium, finite temperature, deconfined plasma 
state corresponds to a static (planar) black hole in the bulk. The Hawking
temperature of this black object is
interpreted as the temperature in the dual \symm. Extending this notion to the 
out-of-equilibrium situation we assume that non-equilibrium plasma states
correspond to geometries with non-static horizons. Such geometries are assumed
to poses an event horizon, but there are reasons to believe \cite{Booth:2005qc,Booth:2010kr,Booth:2011qy} that
the physical notions (such as entropy for instance) should be
associated with {\it apparent horizons}.

This conjecture allows us to extend the notion of entropy to non-equilibrium
states by the Bekenstein-Hawking relation, which with our normalization
translates to 
\begin{equation}
S = \frac{a_{\rm AH}}{\pi}~,
\label{eq:entropy}
\end{equation}
where $a_{\rm AH}$ is apparent horizon area \cite{Booth:2009ct}. By the area
law theorems this quantity is non-decreasing and agrees with thermodynamic definition
for late times. 

In calculations performed here we took $600$ different initial states 
described by randomly generated initial geometries -- with each of these we
associate an initial entropy as defined above. We then evolved these
geometries according to Einstein equations well into the hydrodynamic regime. From the
rules of the holographic correspondence we are able to read of the 
relevant physical observables i.e. energy density $\epsilon(\tau)$ and
entropy $S(\tau)$. An important assumption on the initial conditions
is that $\epsilon(0)\neq 0$, which allows us to normalize the initial effective temperature
as $T(0)=1/\pi$. For more technical details of construction of initial
geometries and solving for the time evolution we refer to the original
paper \cite{Jankowski:2014lna}.

%%%%%%%%%%%%%%%%%%%%%%%%%%%%%%%%%%%%%%%%%%%%%%%%%%%%%%%%%%%%%%%%%%%%%%%%%%%%%%%%%%%%%

\section{Results and conclusions}

In order to present quantitative results on the hydrodynamization process
we need to give it a precise definition. The approach to hydrodynamics can be
observed by monitoring the pressure 
anisotropy
\be
\Delta \equiv \frac{p_T-p_L}{\epsilon} \, .
\eb
It is convenient
to measure time in units of inverse local temperature, that is, to use $w=\tau
T(\tau)$ as a parameter of evolution. The pressure anisotropy can then be
expressed as $\Delta(w)= 6 f(w) - 4$ in terms of the function (first
introduced in \cite{Heller:2011ju}) 
\begin{equation}
f(w) = \frac{\tau}{w}\frac{d w}{d\tau} .
\label{eq:fdef}
\end{equation} 
which for dimensional reasons is independent of $\Lambda$. At large times
(large $w$) this function attains a universal 
(hydrodynamic) form $f_{H}(w)$, 
determined by an infinite set of transport coefficients, 
which up to the third order reads
\begin{equation}
f_H(w) = \frac{2}{3}+ \frac{1}{9\pi w} +
\frac{1-\log 2}{27\pi^2 w^2}+
\frac{15-2\pi^2-45\log 2+24 \log^2 2}{972 \pi^3 w^3}~.
\label{eq:fhydro}
\end{equation}
The beginning of the hydrodynamic stage might now be defined as the value
of $w$ (hence proper time $\tau$) when the difference between the actual
$f(w)$ and the hydrodynamic form $f_H(w)$ is less than some arbitrary small
number; for example
\begin{equation}
|\frac{f_H(w)}{f(w)}-1|<0.05 \, .
\label{eq:crit}
\end{equation}
While this definition involves some arbitrary choice, varying this criterion
within reason leads to no appreciable change in the calculated value of the
hydrodynamization time. 

%%%%%%%%%%%%%%%%%%%%%%%%%%%%%%%%%%%%%%%%%%%%%%%%%%%%%%%%%%%%%%%%%%%%%%%%%
%%
%%   use this format to include an .eps figure into your paper
%%
\begin{figure}[htb]
\begin{center}
\includegraphics[width=0.6\textwidth]{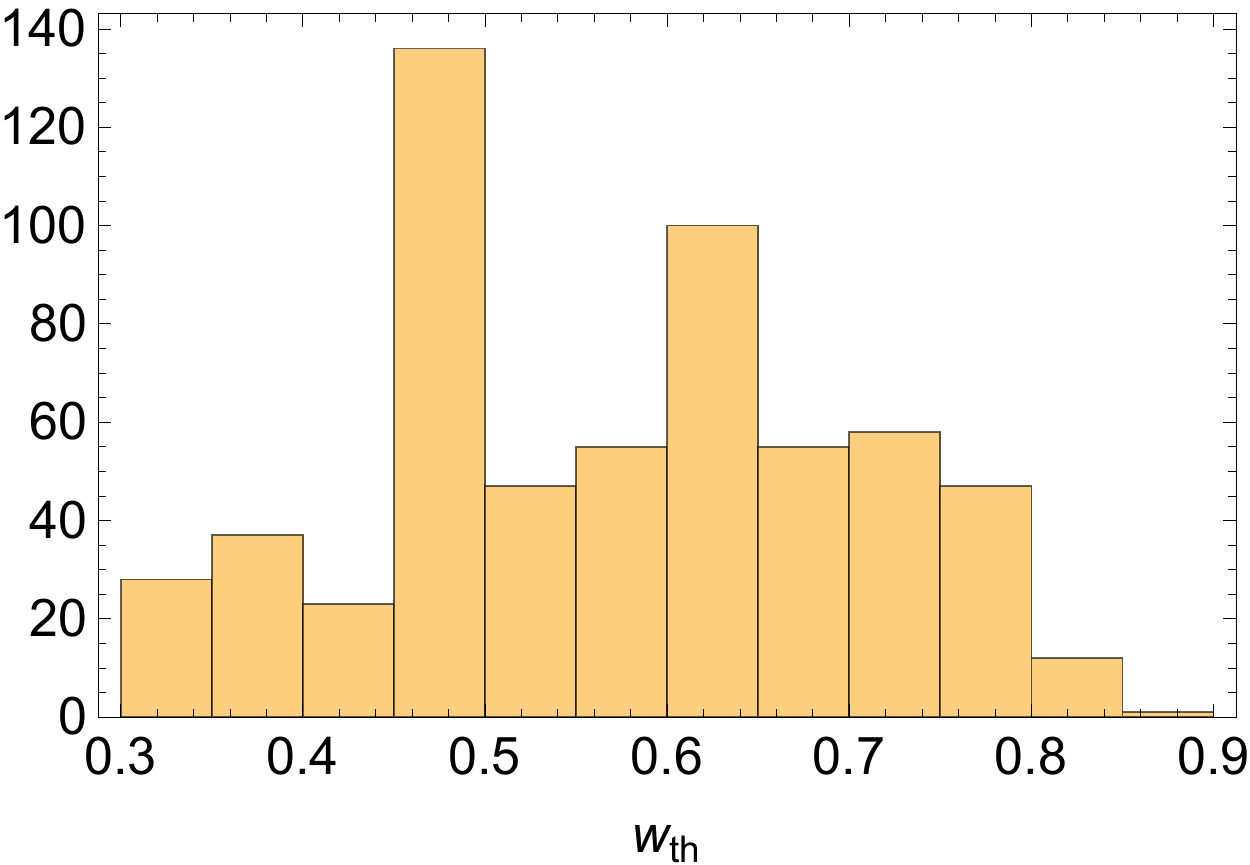} 
\caption{Histogram of hydrodynamization times $w_{\rm th}$ in units of effective
hydrodynamization temperature.}
\label{fig:Hist}
\end{center}
\end{figure}
%
%%%%%%%%%%%%%%%%%%%%%%%%%%%%%%%%%%%%%%%%%%%%%%%%%%%%%%%%%%%%%%%%%%%%%%%%%%%

Figure \ref{fig:Hist} shows a histogram of hydrodynamization times obtained this
way.  Regardless the initial entropy, the hydrodynamization time is of the order
if the inverse hydrodynamization temperature. On the average $w_{\rm av}=0.57$. It
is instructive to compare it to the estimation from the RHIC data; $T=500$~MeV
and $\tau=0.25$~fm/c gives $w_{\rm RHIC}=0.63$ which is very close to
theoretical prediction. The pressure anisotropy at hydrodynamization is found to
be quite high: $\Delta\approx 0.35$. 

The other quantity we focus on here is the energy scale $\Lambda$ which sets
the scale for the 
hydrodynamic cooling. This quantity is obtained by 
fitting the tail of the data to \rf{eq:Th}. 

%%%%%%%%%%%%%%%%%%%%%%%%%%%%%%%%%%%%%%%%%%%%%%%%%%%%%%%%%%%%%%%%%%%%%%%%%
%%
%%   use this format to include an .eps figure into your paper
%%
\begin{figure}[htb]
\begin{center}
\includegraphics[width=0.6\textwidth]{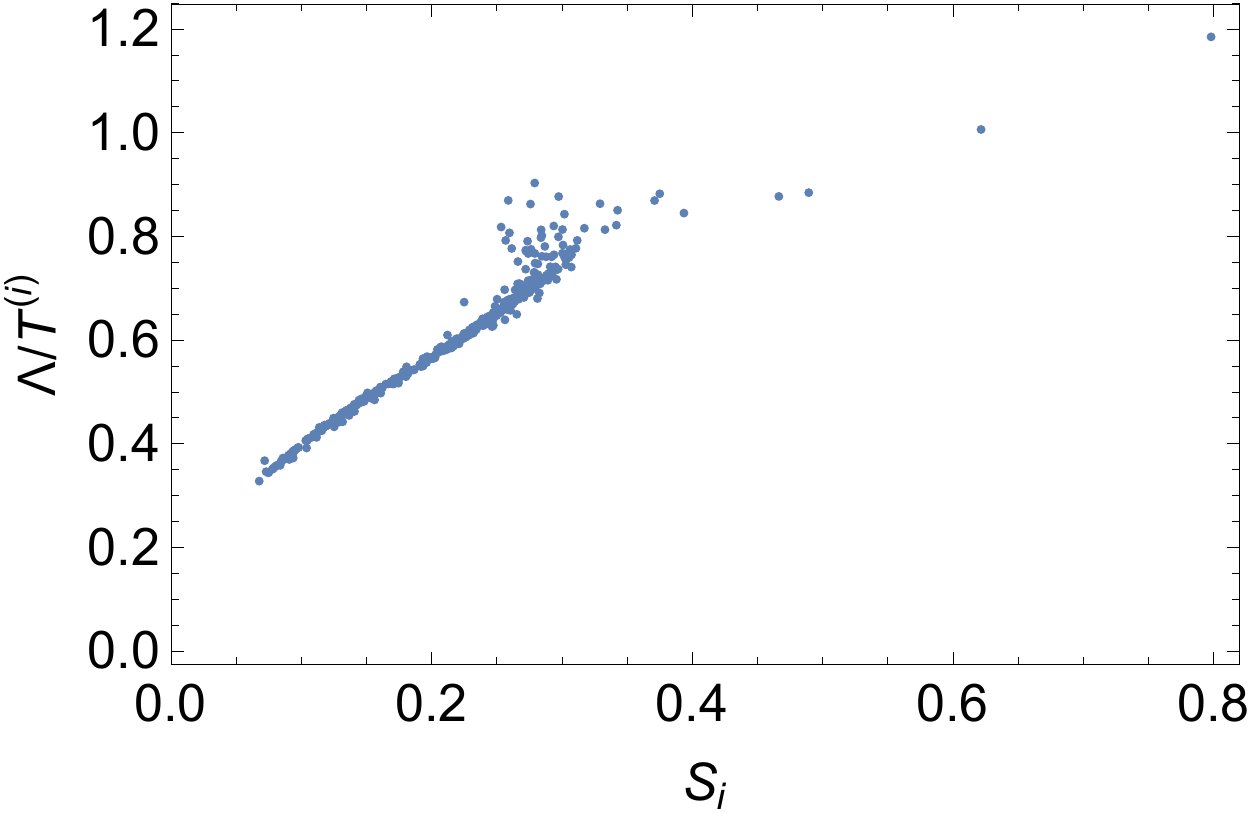}
\caption{Hydrodynamic energy scale $\Lambda$ in units
of initial effective temperature.}
\label{fig:Lambda}
\end{center}
\end{figure}
%
%%%%%%%%%%%%%%%%%%%%%%%%%%%%%%%%%%%%%%%%%%%%%%%%%%%%%%%%%%%%%%%%%%%%%%%%%%%

Results for this scale are shown in figure \ref{fig:Lambda}.
For an intermediate range of entropies below $S\sim0.3$ 
a strong linear correlation is observed. For initial entropies larger than
$0.3$ the correlation is lost and ``chaotic'' behaviour develops. 

In conclusion, through our analysis of the time evolution of a bulk sample of
different non-equilibrium  
initial states we find significant evidence to claim that the early
 hydrodynamiztion of \cite{Heller:2011ju} is a generic process. 
Our analysis supports previous findings that hydrodynamiztion
is different from thermalization in the sense of non-negligible pressure
gradients being present and well described by hydro. It also suggests that
entropy production during the hydrodynamic stage of evolution is not 
  negligible despite the low value of $\eta/s$. 

Our simulations suggest that, at least in 
some regions of initial state parameters, there might exist characteristic 
regularities, reflecting the nature of collective non-hydrodynamic degrees of
freedom \cite{Peschanski:2012cw}.
At this point an important question to what extent is  this correlation 
a consequence of the strong symmetry assumptions imposed and to what extent it
reflects the true nature of the process.

\subsection*{Acknowledgements}

I would like to thank M. Spali{\'n}ski and G. Plewa for collaboration on
this project and S.~Mr{\'o}wczy{\'n}ski  for useful comments. 
I express my thanks to the organizers of the CSQCD IV conference for providing an 
excellent atmosphere which was the basis for inspiring discussions with all participants.
We have greatly benefited from this. This research was supported by a post-doctoral internship grant
No. DEC-2013/08/S/ST2/00547. 
%(JJ) and the Polish National Science Center grant
%2012/07/B/ST2/03794 (MS). 

\end{document}

%% file: econfmacros-csqcd4.tex
\def\beq{\begin{equation}}
\def\eeq#1{\label{#1}\end{equation}}
\def\eeqn{\end{equation}}

\def\beqa{\begin{eqnarray}}
\def\eeqa#1{\label{#1}\end{eqnarray}}
\def\eeqan{\end{eqnarray}}

\let\bar=\overbar

\def\L{{\cal L}}

\def\Dslash{\not{\hbox{\kern-4pt $D$}}}
\def\dslash{\not{\hbox{\kern-2pt $\del$}}}

\def\msb{{\bar{\ssstyle M \kern -1pt S}}}

\usepackage{fancyhdr,graphicx}